\documentclass{article}

\PassOptionsToPackage{numbers, compress}{natbib}
\usepackage[preprint]{neurips_2023}

\usepackage[utf8]{inputenc} % allow utf-8 input
\usepackage[T1]{fontenc}    % use 8-bit T1 fonts
\usepackage{hyperref}       % hyperlinks
\usepackage{url}            % simple URL typesetting
\usepackage{booktabs}       % professional-quality tables
\usepackage{amsfonts}       % blackboard math symbols
\usepackage{nicefrac}       % compact symbols for 1/2, etc.
\usepackage{microtype}      % microtypography
\usepackage{xcolor}         % colors
\usepackage{graphicx} 

\title{Improving International Climate Policy via Mutually Conditional Binding Commitments (Track 3)}

\author{
Jobst Heitzig \\
FutureLab on Game Theory and Networks of Interacting Agents\\ 
Complexity Science Department\\
Potsdam Institute for Climate Impact Research \\
\texttt{jobst.heitzig@pik-potsdam.de}
\And
Jörg Oechssler \\
Alfred Weber Institute for Economics\\
University of Heidelberg
\And
Christoph Pröschel \\
Technical University of Berlin \\
\texttt{c.proeschel@campus.tu-berlin.de}
\And
Niranjana Ragavan
\And
Richie YatLong Lo \\
Dyson Robot Learning Lab \\
\texttt{richie.lo@dyson.com}
}

\begin{document}

\maketitle

\begin{abstract}
  This paper proposes enhancements to the RICE-N simulation and multi-agent reinforcement learning framework to improve the realism of international climate policy negotiations. Acknowledging the framework's value, we highlight the necessity of significant enhancements to address the diverse array of factors in modeling climate negotiations. Building upon our previous work on the "Conditional Commitments Mechanism" (CCF mechanism) we discuss ways to bridge the gap between simulation and reality. We suggest the inclusion of a recommender or planner agent to enhance coordination, address the Real2Sim gap by incorporating social factors and non-party stakeholder sub-agents, and propose enhancements to the underlying Reinforcement Learning solution algorithm. These proposed improvements aim to advance the evaluation and formulation of negotiation protocols for more effective international climate policy decision-making in Rice-N. However, further experimentation and testing are required to determine the implications and effectiveness of these suggestions.
\end{abstract}

\section{Introduction}

The current version of the RICE-N simulation and the multi-agent reinforcement learning framework provides valuable methods to evaluate and formulate negotiation protocols among the regions for effective climate policy negotiations. However, as with every other model, there are some aspects of the framework that needs significant improvement for a realistic simulation that addresses and takes into account most of the factors that directly or indirectly influence international climate decisions. 

In Track 2, we have discussed our proposed solution, the “Conditional Commitments Mechanism” (CCF mechanism)\cite{heitzigSSRN}. A possible real-world implementation of the CCF mechanism would be via unilateral legislation whereby an individual country passes a bill that (i) specifies a CCF encoding offers and conditions, and (ii) bindingly commits that country to comply with the largest action profile that is feasible (in the above sense) given all CCFs that are specified in similar bills that are put into force by other countries. 

In this essay, we are aiming to discuss some of the improvements to the RICE-N framework and reinforcement learning agents that would be effective in narrowing the gap between the simulation and the real world. 

\section{Inclusion of Recommender or planner agent}

In the current scenario, the RICE-N framework has only agents for countries (world regions) and does not allow for additional agents that might represent non-country actors such as an international agency that could guide the negotiations by, e.g., making proposals. Therefore in our simulations, we had to assume that there is no current central entity for leading the negotiation. It is true that the presence of a central entity that acts in enforcing the agreements among the agents in the real world is practically difficult, as the countries need to accept the entity as trusted and also it would be uncertain where the central entity would draw the power from. Moreover, this would reduce the autonomy of the decisions and also lead to adverse effects if the entity's interests are not in alignment with the agents. Still, allowing for the specification of additional agents who can send signals during negotiations can improve coordination among the regional agents.

There are some factors that should be considered (which are discussed later in this essay) during the complex interaction among the agents, which decide the feasibility of negotiations. In this case, there is a need for the common framework or mediator which aids the agents via some of the following (not limited to) mechanisms:
\begin{itemize}
    \item Recommendations or proposals for effective negotiations (the implementation of this system could be either made by training the mediator agent independently of the agents with the help of historical information of the agents’ negotiation or training the mediator agent,  as an observer along with the other agents. It is still needed to be experimented on how this training shapes the mediator agent’s belief system and how to ensure a neutral perspective, or whether it would be effective if the mediator makes randomized recommendations which might help in convergence to the equilibrium faster.
    \item Verification of the truthfulness and completeness of the publicly shared information by the agents.\cite{digiovanni2022commitment}
    \item Verification or evaluating the degree to which the agent could comply with the binding agreement, as a metric of feasibility.
\end{itemize}
The central entity need not intervene in the negotiation process or enforce the negotiation. While it could be used as an aid which provides proposal, validation and verification mechanisms in the framework.

\section{Addressing the Real2Sim variation}

In the current framework, the agents are assumed to be boundedly rational in the sense that they are not planning their policies via some form of game-theoretic analysis of the game but have to learn well-performing policies via some form of reinforcement learning. Also, more complex interactions and other lurking factors, e.g. the presence of domestic veto players such as powerful lobby groups or factions in parliament, have been excluded. To reduce the gap between real and simulated environments, the socio-economic and cultural factors of different countries, that could be measured quantitatively to some extent, should be included. Some features that could be added to the simulation are

\begin{itemize}
    \item Social influence as a measure in their belief system, to prevent the agent to be overly optimistic about its economic or climate outcomes, or to aid the agents to visualize their stand among other agents, which could be used as a penalty for their non-compliance.
    \item The inclusion of sub-agents or non-party stakeholders along with their respective interests and climate economic outcomes \cite{mathias2020grounding}.
\end{itemize}

Even though the binding agreements have been made successfully among the regions, we are assuming that the region would ensure compliance internally. While in the complex real world, cultural differences, domestic politics or civil unrest, etc., could limit the capacity of the region to successfully implement its agreement. In this case, mere agreements would not be of much help. There has to be a set of validation or prediction metrics to evaluate whether the particular region is capable of abiding by the agreement. 

Let’s say A has agreed to the mitigation rate of 12\%. The other agents or a verification strategy could evaluate the probability that country A could successfully abide by the agreement, using the parameters that indicate agent A's internal situations. Based on the metric, extended help or additional measures such as adjusting the liabilities could be done. This could suggest a more practical and feasible implementation of international negotiations.

\subsubsection*{Observability of features}

In the current RL framework, There are some public variables of the agents that are visible to other agents in their observation space, such as capital, mitigation rate, and others. While in the real-world scenario, some of the assumed variables in the framework might not be publicly observable, or only approximately and with considerable delay. Thus, further careful curation of these variables should be needed. This could widely alter the belief system of the agents that are getting trained.

\subsubsection*{Mutual transparency}

In this case, it has been assumed that all the information (public variables) that is shared among the agents is completely and truthfully reported by the respective agents. While in the real world, this private information could be subjected to an altercation or misreporting by the agents. The way these variables are represented might affect the belief system of other agents and their respective actions. Thus, there has to be a verification mechanism for mutual transparency among the agents. This could be enforced via the central recommendation engine (mentioned above) or presenting the mutual transparency (and related monitoring mechanisms) itself as a binding agreement (while the latter could cause more complex interactions in the system).

\subsubsection*{Including negotiation component using natural language in the framework }

While the current framework represents the formal bargaining case, in the real world, the use of natural language and diplomatic communication plays an important role in effectively mitigating their needs and their actions. Thus, we suggest that the need for a Negotiation component could be developed in future to aid the current framework \cite{meta2022human}. This could also help in communicating this framework successfully to lawmakers. Yet there are some considerations here:

\begin{itemize}
    \item If an agent makes a conditional commitment or binding agreement, will agents successfully understand each other’s adherence based on the natural language?
    \item Should there be a penalty or reward that is provided as feedback to the agent that evaluates the effectiveness of its previous exchange of communications? 
\end{itemize}

\section{Suggested improvements in RL algorithms}

It was somewhat unclear to us whether the RL algorithms supplied to participants (based on the A2C algorithm) can be expected to converge to a Nash equilibrium (NE) or some refinement of NE (such as subgame-perfect or strong equilibrium), or at least to a low-regret state of the negotiation game in the first place, and if so to which particular equilibrium if there are many (as is usually the case in complex games like this).

From comparing the used RL approach with established alternative approaches to find equilibria used in environmental economics (e.g., optimization with welfare weights using gradient descent in action space), we suspect that the used RL agents might not be reliably able to find Pareto-efficient and collectively rational equilibria for at least two independent reasons.

\subsubsection*{No discretization of continuous action space}

For one, the RL agents use a randomized policy over a discretized action space rather than a deterministic policy over the actual continuous action space. 
Learning via gradient descent thus cannot exploit the topological structure of the action space. 
Rather than shifting actions smoothly around in action space, they shift probability mass from points in action space to possibly quite far away regions in action space. 
But given the largely approximately convex public-good structure of the relevant utility functions, it is not to be expected that Pareto-efficient NEs are mixed rather than pure strategies, and that a classical gradient descent in action space (rather than in probability space) should converge well. 
we hence suggest dropping the action discretization and use algorithms such as the continuous-action version of A2C \cite{continuous}.

\subsubsection*{Meta-learning MARL and anticipation}

Independently of this, the fact that the learning algorithm for the RL agent representing country $i$ is treating the RL agents representing the other countries as part of the environment makes the environment non-stationary since all RL agents are trained simultaneously and thus have changing policies.
Since convergence theorems usually have to assume stationarity, one can expect weak convergence because of this non-stationarity. 
This is not surprising given the longstanding insights from the early theory of learning in games that showed that already simple best-response dynamics can lead to oscillations rather than convergence since each player is ``chasing a moving target''. 
To overcome this, we suggest using state-of-the-art Multi-Agent Reinforcement Learning (MARL) algorithms that treat the environment and other agents differently, such as the ``meta-learning'' approach in \cite{mlearn}.

We believe that particularly designs in which an RL agent anticipates the other RL agents' next learning steps would be very conducive to convergence in a negotiation context. 
This is because if $i$ switches in a learning step from making a bad offer to making a good offer, this will only pay off after the other players have switched in their next learning step from not accepting the offer to accepting the offer.

\section{Conclusion}

In this essay, we have suggested a few of the improvements and extensions that could be included in the original framework which could help in developing a more realistic version of climate negotiation protocols. However, the implications and effectiveness of these improvements could be decided in an absolute way only after implementing and testing them in a systematic setting. These could also be affected by how we interpret the climate negotiation problem (as cooperative, non-cooperative, or mixed one)and its objective function.  

\subsection*{Author contributions}
N.R. drafted the paper. J.H., C.P., and N.R. edited and finalized the paper.

{\small\sffamily
\bibliographystyle{IEEEtran}
\bibliography{refs} %{REFERENCES}
}

\end{document}